\documentclass[aps,prb,twocolumn,groupedaddress,floatfix]{revtex4-2}

\usepackage{amsmath,amssymb,graphicx,hyperref}

\usepackage{xcolor}
\usepackage{hyperref}
\hypersetup{
colorlinks,
linkcolor={red!50!black},
citecolor={blue!80!black},
urlcolor={blue!80!black}
}

\renewcommand{\eqref}[1]{Eq.~(\ref{#1})} 
\providecommand{\Em}{E_{-}}
\providecommand{\Ep}{E_{+}}
\renewcommand{\vec}[1]{\mathbf{#1}}
\begin{document}


\title{Plasma frequency in doped highly mismatched alloys}


\author{Hassan Allami}
\affiliation{Department of Physics, University of Ottawa, Ottawa, ON K1N 6N5, Canada}
\author{Jacob J. Krich}
\affiliation{Department of Physics, University of Ottawa, Ottawa, ON K1N 6N5, Canada}
\affiliation{School of Electrical Engineering and Computer Science, University of Ottawa, Ottawa, ON K1N 6N5, Canada}



\begin{abstract}
Highly mismatched alloys (HMAs) have band structures strongly modified due to the introduction of the alloying element. We consider HMAs where the isolated state of the alloying element is near the host conduction band, which causes the conduction band to split into two bands. We determine the bulk plasma frequency when the lower-energy band is partially occupied, as by doping, using a semi-analytical method based on a disorder-averaged Green's function. We include the nontrivial effects of interband transitions to the higher-energy band, which limit the plasma frequency to be less than an effective band gap. We show that the distribution of states in the split bands causes plasmons in HMAs to behave differently than plasmons in standard metals and semiconductors. The effective mass of the lower split band $m^*$ changes with alloy fraction, and we find that the plasmon frequency with small carrier concentration $n$ scales with $\sqrt{n}/m^*$ rather than the $\sqrt{n/m^*}$ that is expected in standard materials.
We suggest experiments to observe these phenomena.
Considering the typical range of material parameters in this group of alloys and taking a realistic example, we suggest that HMAs can serve as highly tunable low-frequency plasmonic materials.
\end{abstract}


\maketitle
\section{Introduction \label{sec:intro}}

\begin{figure}[t]
	\includegraphics[width=\columnwidth]{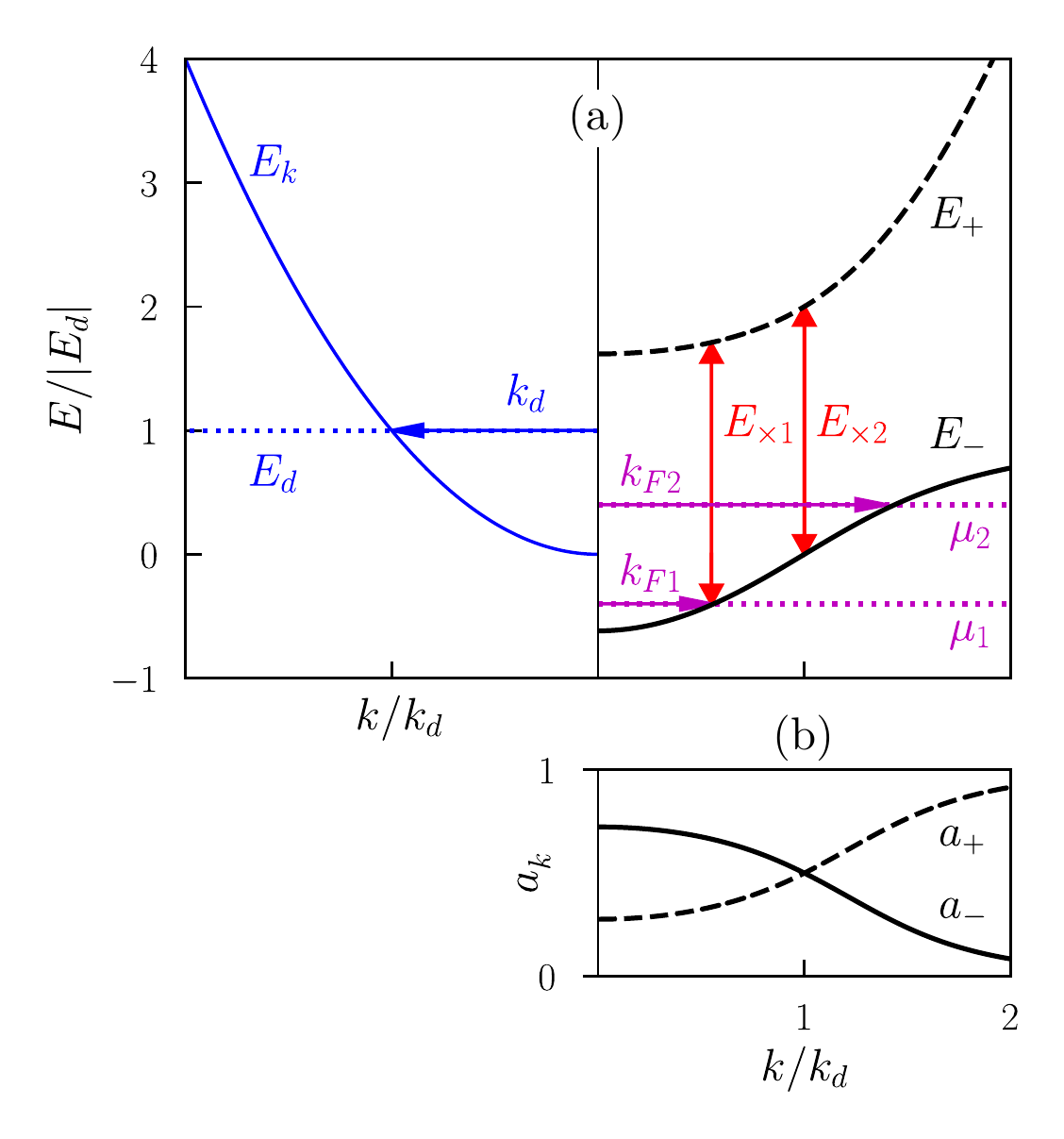}
	\caption{(a) Band structure of a generic HMA with $E_d > 0$ and (b) corresponding weighting factors .
	(a-left) Impurity level, $E_d$, and CB of the host material, which is taken to be parabolic, $E_{\vec{k}} = k^2/2m$.
	The $k$-axes are normalized to $k_d$, defined in \eqref{eq:kd} and also indicated.
	The energy axis is normalized to $|E_d|$.
	(a-right) The corresponding two split bands of BAC, $E_\pm$, according to \eqref{eq:Epm}, for $V^2x = E_d^2$.
	Also, for the two different cases of \eqref{eq:E-cross-p}, two different levels of filling are shown by $\mu$'s and $k_F$'s, with their corresponding inter-band energy gaps, $E_{\times1}$, $E_{\times2}$.
	(b) Weighting factors $a_\pm$ for this realization, from \eqref{eq:weights}.}	
	\label{fig:bands-weights-p}
\end{figure}

\begin{figure}[ht]
	\includegraphics[width=\columnwidth]{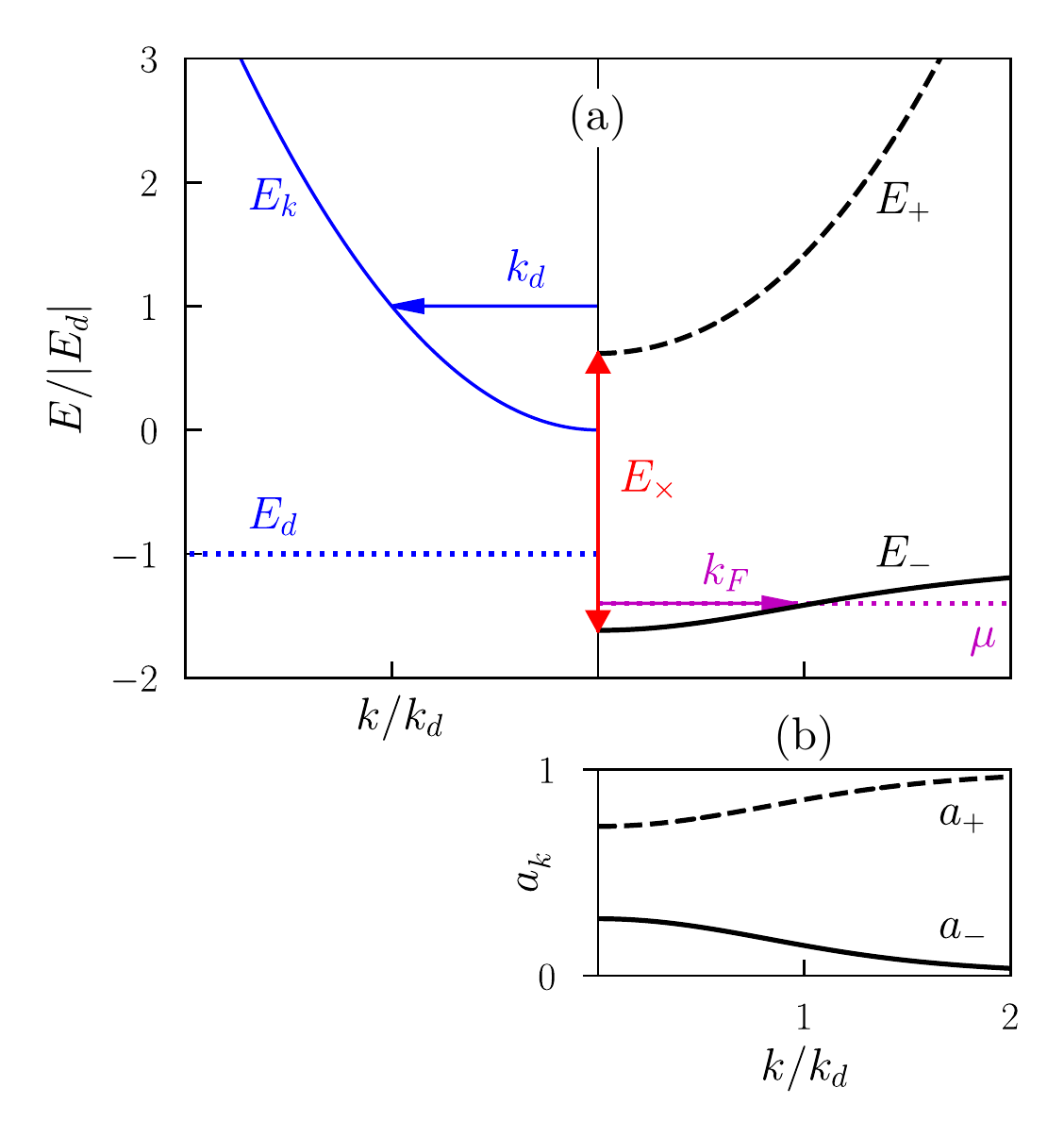}
	\caption{(a) Band structure of a generic HMA with $E_d<0$ and (b) corresponding weighting factors.
	(a-left) Impurity level, $E_d$, and CB of the host material, which is taken to be parabolic.
	The $k$-axes are normalized to $k_d$, defined in \eqref{eq:kd}, and also indicated.
	(a-right) The corresponding two split bands of BAC, $E_\pm$, according to  \eqref{eq:Epm}, for $V^2x = E_d^2$.
	A chemical potential, $\mu$, and its corresponding $k_F$ are also shown,
	along with the inter-band energy gap, $E_\times$, which is independent of filling when $E_d<0$.
	(b) Weighting factors $a_\pm$ for this realization, from \eqref{eq:weights}.}
	\label{fig:bands-weights-n}
\end{figure}

Highly mismatched alloys (HMA) are semiconductor compounds in which atoms of significantly different electronegativity substitute the host atoms.
They are characterized by an unusually strong variation in their fundamental band gap upon the introduction of a small fraction of substituting elements.
The band anticrossing (BAC) model was proposed to explain this behavior \cite{bac-original}.
In the BAC, the highly mismatched substitute atoms form localized states with energy $E_d$ near the continuum of extended states of the host material, $E_\vec{k}$.
The localized states strongly couple to the host's extended states leading to the formation of two split bands, $E_\pm$ (see Figs.~\ref{fig:bands-weights-p}a and \ref{fig:bands-weights-n}a).

The dramatic band gap drop, first observed in the 1990s~\cite{GaAsN-bandgap-drop-92}, soon found applications such as developing quantum well laser diodes~\cite{GaN-led} and high efficiency multi-junction solar cells~\cite{4th-multi-SC}.
Later, building on the BAC model of two split bands in HMAs, they were used to implement intermediate band solar cells~\cite{Lopez11, Ahsan12,Welna17,Tanaka17a,Heyman17,Zelazna18,Heyman18}, which is another scheme for harvesting sub-band-gap photons~\cite{luque-marti-1997}.
A collection of recent developments in the study of this class of semiconductor alloys can be found in a Special Topic issue of {\em Journal of Applied Physics}~\cite{HMA-special-topic}.

The plasmonic properties of the split bands in HMAs have not been studied. As with any band, the presence of mobile charge carriers  results in a negative dielectric function below and near the resonant plasma frequency of the medium. The interface of such a medium with a surrounding dielectric supports localized surface modes that are the basis of plasmonic phenomena such as the subwavelength walking waves of surface plasmon polaritons~\cite{SPP-theory} or the standing modes of localized surface plasmons on nanoparticles~\cite{nanoplasmonics}. The operating frequency range of these modes is determined by the bulk plasma frequency $\omega_p$ of the system. 
While the classic plasmonic metals such as  gold and silver have  operating frequencies in the visible and near-infrared range, there has been a search for alternative plasmonic materials that operate in lower frequency ranges, such as terahertz and mid-infrared, which can offer technological advantages such as reducing the size of electronic devices operating in these ranges \cite{beyond-Au-Ag}.

While doping of HMA bands can be a challenge~\cite{cl-dope-ZnTeO-19}, mobile charge in the  $\Em$ band can provide tunable plasmonic effects. We show that both the origins of that band and the close proximity to the $\Ep$ band  make these properties different from standard doped semiconductor plasmonics~\cite{doped-semiconductor-plasmonics}.
As with doped semiconductors, the carrier density in such a system is low compared to noble metals, giving a lower frequency range of plasmonic operation.  The large tunability of both band gaps and doping in HMAs makes them appealing platforms for development of plasmonic structures in the mid-infrared regime. These plasma oscillations may also be important for recombination in HMA-based intermediate band solar cells. 
Moreover, since a gap separates the $\Em$  band from the $\Ep$ band, proper tuning could allow for minimizing loss, which is a crucial favorable feature for plasmonic applications \cite{lossless-matal}.

In this work, we study the long-wavelength limit of bulk plasmons of a model for HMAs.
In particular, we focus on the important role of state distribution, which is beyond the scope of the simpler BAC model.
In Sec.~\ref{sec:chi} we formulate the calculation of the density susceptibility of the system in the long-wavelength limit, which is required for finding $\omega_p$ of the system.
Next, in Sec.~\ref{sec:wp} we set up an equation for $\omega_p$ of the system and analyze the behavior of its solution, emphasizing the important qualitative features.
Finally in Sec.~\ref{sec:conclusion}, we provide numerical values for $\omega_p$ for typical realistic parameters of previously studied HMAs
and conclude by commenting on experimental methods for observing the predicted phenomena.

\section{Susceptibility of HMA bands \label{sec:chi}}

In this section, we construct a formulation for calculating the susceptibility that is needed to find the plasma frequency of a HMA system.
First we show how the state distribution is described by the spectral density of the system, and then we use it in the calculation of susceptibility $\chi$.
We show how the formulation simplifies in the long-wavelength limit and derive the limiting form to be used for our semi-analytical calculations.

\subsection{The spectral density of HMAs \label{sec:spectral}}

The BAC model considers only the localized impurity levels with energy $E_d$ and one set of extended states of the host material, $E_\vec{k}$, which we consider to be the conduction band (CB) of the host material. 
BAC model introduces a $2\times2$ Hamiltonian at each wavevector $\bf k$~\cite{bac-original}
\begin{equation}
	H_{\rm BAC} = 
	\begin{bmatrix}
	E_\vec{k}&V\sqrt{x}\\
	V\sqrt{x}&E_d
	\end{bmatrix},
	\label{eq:Hbac}
\end{equation}
where $V$ is the average coupling between localized and extended levels, and $x$ is the fraction of impurity atoms. This matrix can be diagonalized to give two split bands with energies
\begin{equation}
	E_\vec{k}^\pm = \frac{1}{2}\left(E_\vec{k} + E_d \pm 
	\sqrt{(E_\vec{k} - E_d)^2 + 4V^2 x}\right).
	\label{eq:Epm}
\end{equation}
Measuring energy from the bottom of the CB of the host material, we identify two cases: when the impurity level is inside the CB, $E_d > 0$, as in Fig.~\ref{fig:bands-weights-p}a, and when the impurity level is below the band edge of the CB, $E_d < 0$, as in Fig.~\ref{fig:bands-weights-n}a.

While this model provides a good description of the energy spectrum of HMAs, it does not preserve the total number of states, as it implies two $\vec{k}$-states for each extended $\vec{k}$-state of the host material, which is incorrect for small $x$.
Early in the development of BAC model, Wu et al., based on Anderson's impurity model~\cite{anderson1961}, proposed an average Green's function
\begin{equation}
	G(E,\vec{k}) = \left[E - E_{\vec{k}} -
	\frac{V^2 x}{E -E_d + i\Gamma}\right]^{-1},
	\label{eq:the-GF}
\end{equation}
which does not suffer from this issue \cite{bac-gf}.
Here, $\vec{k}$ is still a good quantum number in the ensemble average sense \cite{elliott-cpa-review}.
In this Green's function, $\Gamma$ is a broadening factor given by $\pi\beta V^2\rho_0(E_d)$, where $\rho_0(E_d)$, which has dimensions of inverse energy, is the unperturbed density of states in a unit cell at the energy of the defect level, and $\beta$ is a number of the order of 1.

As we will show, the distribution of states with energy plays an important role in the dynamics of bulk plasmons in HMAs.
The spectral density of a Green's function properly describes the distribution of states in a system.
It is well known that the spectral density (or spectral function) of a retarded Green's function is given by $A_\vec{k}(E) = -{\rm Im}[G(E, \vec{k})]/\pi$, from which the HMA spectral function is
\begin{equation}
	A_\vec{k}(E) = \frac{1}{\pi}\frac{\Gamma V^2 x}
	{[(E - E_d)(E-E_\vec{k}) - V^2x]^2 +
	\Gamma^2(E - E_\vec{k})^2}.
	\label{eq:A}
\end{equation}
One can check that for each $\bf k$, the integral of $A_\vec{k}(E)$ in~\eqref{eq:A} over energy equals 1; $A_\vec{k}(E)$ describes how each $\bf k$ is distributed among all energies due to hybridization of the localized and propagating states. 
Note that when the broadening factor $\Gamma$ is sufficiently small, the spectral density in \eqref{eq:A} has two sharp peaks near the split bands of BAC model, $E_\vec{k}^{\pm}$ in~\eqref{eq:Epm}, as they are the roots of the square bracket in the denominator of $A_\vec{k}(E)$. This relationship shows how the Green's function in \eqref{eq:the-GF} contains the BAC model.

The broadening factor $\Gamma$ is naturally much smaller than $V$, most obviously when $E_d<0$, in which case $\Gamma = 0$.
Consider a parabolic host CB and define $k_d$ such that $E_{k_d}=|E_d|$, as illustrated in Figs.~\ref{fig:bands-weights-p}a and \ref{fig:bands-weights-n}a, giving
\begin{equation}
	k_d = \frac{\sqrt{2m |E_d|}}{\hbar}.
	\label{eq:kd}
\end{equation}
Then in the $E_d > 0$ case, approximating the size of the unit cell by $k_{\rm BZ}^{-3}$, where $k_{\rm BZ}$ is the length of the Brillouin zone,
we have $\rho_0(E_d)~\approx~mk_d k_{\rm BZ}^{-3}/(\hbar \pi)^2$.
We also introduce an equivalent momentum for the coupling factor, $k_V = \sqrt{2mV}/\hbar$,
and find $\Gamma/V \approx \beta k_d k_V^2/(2\pi k_{\rm BZ}^3)$.
Then, since $k_d \ll k_{\rm BZ}$ and $k_V\lesssim k_{\rm BZ}$ (since $V$ is of the order of eV), we still have $\Gamma~\ll~V$.
For instance, Heyman et al.\ use $\Gamma = 10^{-3} V$ in modeling their measurements of samples of GaN$_x$P$_y$As$_{1-y-x}$ (see Table I in Ref.~\onlinecite{heyman-17}).

On this basis, we consider the limit
$\Gamma\to 0$, where the spectral density in \eqref{eq:A} turns into two weighted delta functions centered at the split bands of BAC model, $E_\vec{k}^\pm$ in ~\eqref{eq:Epm}, indicating the share of each one at a given $\bf k$
\begin{equation}
	A_\vec{k}(E) = 
	a_\vec{k}^{-}\delta(E - E_\vec{k}^-)+a_\vec{k}^{+}\delta(E - E_\vec{k}^+),
	\label{eq:sharp-A}
\end{equation}
where the weighting factors are given by
\begin{equation}
	a_\vec{k}^\pm = \frac{V^2 x}{|E^\pm_\vec{k} - E_\vec{k}|(E_\vec{k}^+-E_\vec{k}^-)}.
	\label{eq:weights}
\end{equation}
One can check that at any given $\bf k$ the expressions in \eqref{eq:weights} satisfy, $a_\vec{k}^- + a_\vec{k}^+ = 1$, as expected.

These weighting factors $a_\vec{k}^\pm$ are shown in Figs.~\ref{fig:bands-weights-p}b and \ref{fig:bands-weights-n}b.
Notice that $a_\vec{k}^-$ has its maximum at $k = 0$ and that in the case of $E_d > 0$, it is larger than $a_k^+$ for $k < k_d$, while in the case of $E_d < 0$, it is always smaller than $a_k^+$.

\subsection{Long-wavelength limit of the susceptibility \label{sec:longwavelength}}
We use the Green's function of \eqref{eq:the-GF} and the spectral density in the limit of small $\Gamma$, \eqref{eq:sharp-A}, to construct the susceptibility, dielectric function, and plasma frequency of an HMA described by \eqref{eq:the-GF} at zero temperature. We consider the case where doping ensures the chemical potential is in the $E_-$ band. We assume the  valence band lies far below the chemical potential and can be ignored. In this model, $E_\vec{k}$, $E_d$, and $V$ are fixed parameters of the material.
The alloy fraction, $x$, can be tuned, and doping controls chemical potential $\mu$. Since the model is isotropic, we parametrize the chemical potential using the Fermi momentum, $k_F$.
Two examples of $E_\pm$ with $E_d$ above and below zero are shown in Figs.~\ref{fig:bands-weights-p}~and~\ref{fig:bands-weights-n}. Also shown is the effective inter-band energy gap, $E_\times$, which is the smallest difference in energy between an $\Ep$ state and an occupied $\Em$ state at the same $\vec{k}$
\begin{align}
	E_\times=\min_{k<k_F} \left(E^+_\vec{k} - E^-_\vec{k}\right).
	\label{eq:E-cross}
\end{align}
As shown in Fig.~\ref{fig:bands-weights-n}, when $E_d < 0$ we always have $E_\times = \sqrt{E_d^2 + 4V^2x}$, regardless of the filling of $\Em$, as the minimum gap occurs at $k = 0$. For $E_d > 0$ however, as in Fig.~\ref{fig:bands-weights-p}, we have
\begin{equation}
	E_\times = \begin{cases}
		\sqrt{(E_d - E_{k_F})^2 + 4V^2x} & k_F < k_d \\
		2V\sqrt{x} & k_F \geq k_d
		\end{cases},
	\label{eq:E-cross-p}
\end{equation}
since the minimum value of $\left(E^+_\vec{k} - E^-_\vec{k}\right)$ occurs at $k = k_d$.

Bulk plasmons of this system are rooted in the collective density oscillations of free electrons in the $\Em$~band.
This resonant mode can be seen as an instability when the real part of the dielectric function $\epsilon$ equals zero.
The plasma frequency $\omega_p$ of the system is the frequency associated to the long-wavelength limit of this resonance.
Therefore, to compute $\omega_p$ we need a suitable expression for $\epsilon$.
The well-known random phase approximation relation, which connects density susceptibility of a system $\chi$ to $\epsilon$, is particularly suitable for a case where the main dynamics are due to mobile electrons~\cite{giuliani_vignale} 
\begin{equation}
	\epsilon({\bf q}, \omega) = 1 - \frac{4\pi e^2}{q^2}\chi({\bf q}, \omega),
	\label{eq:epsilon-RPA}
\end{equation}
where $e$ is the electron charge and $\bf q$ and $\omega$ are the external wavevector and frequency, respectively. Since we seek $\epsilon(0,\omega_p)=0$, this form reduces the problem to finding the long-wavelength limit of $\chi$.

\begin{figure}[ht]
	\includegraphics[width=0.5\columnwidth]{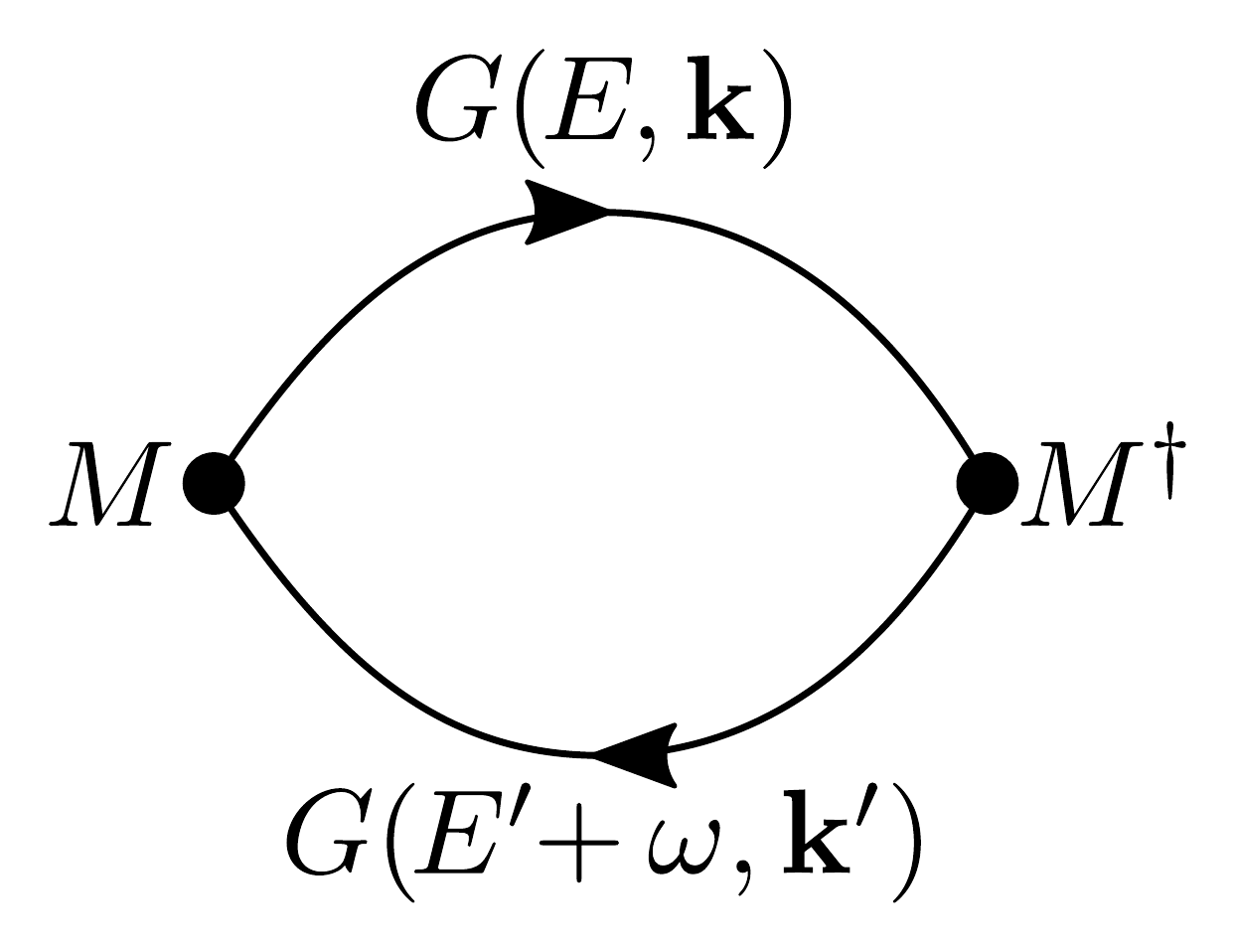}
	\caption{The bubble diagram representing the density susceptibility, $\chi({\bf q}, \omega)$.
	\label{fig:bubble}}
\end{figure}

Generally, $\chi$ can be expressed by a bubble diagram as in Fig.~\ref{fig:bubble}, through the Green's function of propagating particles in the system, $G(E, \vec{k})$ in \eqref{eq:the-GF}, and the matrix element,
$M = \langle \vec{k}, E |e^{i{\bf q\cdot r}}| {\bf k'}, E'\rangle$,
which is on the vertex.
Expressing the $G$ through its spectral density using Lehmann representation \cite{lehmann1954},
the bubble diagram represents the integral
\begin{widetext}
\begin{equation}
	\chi({\bf q}, \omega) = \frac{2}{\mathcal{V}}\int dE dE' 
	\sum_{\bf kk'}
	\frac{A_\vec{k}(E)A_{\bf k'}(E')
	|\langle \vec{k}, E|e^{i{\bf q \cdot r}}|\vec{k}', E'\rangle|^2
	\left[f(E' - \mu) - f(E-\mu)\right]}
	{E'-E + \omega + i\eta},
	\label{eq:chi-integral}
\end{equation}
\end{widetext}
where $\mathcal{V}$ is the volume of the system, $f$ is the Fermi distribution, and $\eta$ is an infinitesimal positive energy.
In writing \eqref{eq:chi-integral}, $\hbar$ is taken to be 1 and
the factor of 2  accounts for spin degeneracy.
In what follows, the denominator is always off resonance, so we set $\eta=0$.

Using the sharp spectral density of \eqref{eq:sharp-A}, the $E$ and $E'$ integrals of \eqref{eq:chi-integral} become trivial.
In this case the bubble generates four separate terms, corresponding to four possible pairings of $a_\vec{k}^+$ and $a_\vec{k}^-$. Combining the cross terms together, we obtain three distinct contributions to the susceptibility
\begin{equation}
	\chi({\bf q}, \omega) = \chi_-({\bf q}, \omega)+\chi_+({\bf q}, \omega)+\chi_\times({\bf q}, \omega).
	\label{eq:3chies}
\end{equation}
After standard manipulations to shift the origin of $\bf k$, these terms can be written as
\begin{widetext}
\begin{align}
	\chi_\pm({\bf q}, \omega) &= \frac{4}{\mathcal{V}}\sum_{\bf kk'}
	\frac{a_{\bf k'}^\pm a_\vec{k}^\pm|\langle{\bf k'},\pm|e^{i{\bf q\cdot r}}|\vec{k},\pm\rangle|^2
	(E_{\bf k'}^\pm - E_\vec{k}^\pm) f(E_\vec{k}^\pm - \mu)}
	{\omega^2 - (E_{\bf k'}^\pm - E_\vec{k}^\pm)^2},\label{eq:Chi-pm}\\
	\chi_\times({\bf q}, \omega) &= \frac{4}{\mathcal{V}}\sum_{\bf kk'}
	\frac{a_{\bf k'}^+ a_\vec{k}^-|\langle{\bf k'},+|e^{i{\bf q\cdot r}}|\vec{k},-\rangle|^2
	(E_{\bf k'}^+ - E_\vec{k}^-) [f(E_\vec{k}^- - \mu)-f(E_{\bf k'}^+ - \mu)]}
	{\omega^2 - (E_{\bf k'}^+ - E_\vec{k}^-)^2},
	\label{eq:Chi-cross}
\end{align}
\end{widetext}
where $| \bf k, \pm\rangle$ is the state with wavevector $\bf k$ in the $\Ep$ or $\Em$ band. The $\chi_\pm$ terms correspond to intraband transitions, while $\chi_\times$ corresponds to interband transitions.
We study the case where $\mu$ is in the $E_\vec{k}^-$ band,
so $f(E_\vec{k}^+ - \mu)=0$ at low temperature, and $\chi_+$ can be neglected.

To find $\omega_p$, we need the $\vec{q}\to 0$ limit of \eqref{eq:3chies}. 
In Appendix~\ref{sec:matrix-element}, we use a tight-binding model to argue that the matrix element
$\langle \vec{k}, \pm |e^{i{\bf q\cdot r}}| {\bf k'}, \pm\rangle$ ensures momentum conservation, $\vec{k}' = \vec{k} + \vec{q}$.
We also consider the leading order terms as $\vec{q}\to0$ and argue that for intraband transitions the leading term is simply 1, while in the case of interband transitions 
\begin{equation}
	\lim_{q\to 0} |\langle \vec{k}, - |e^{i{\bf q\cdot r}}| {\bf k'}, +\rangle|^2 \approx q^2l^2\delta_{\bf k',k+q}
	\label{eq:cross-me}
\end{equation}
for some length scale $l$.  In principle, $l$ could be $\vec{k}$-dependent, but for simplicity we consider the case where $l$ is constant.
In a tight-binding framework, $l$ is expected to be of the order of the lattice constant. We estimate the size of $l$ by showing that it is related to the matrix element in the interband absorption coefficient. 
We thus make an order of magnitude estimation of $l$ from transient absorption measurements on some HMAs~\cite{heyman-17}. That analysis is consistent with $l$ being of the order of a lattice constant for the host crystal. 

Putting these results from Appendix~\ref{sec:matrix-element} into Eqs.~(\ref{eq:Chi-pm},\ref{eq:Chi-cross}), for small $\vec{q}$,
\begin{widetext}
\begin{align}
	\chi_-(\bf q, \omega) &= \frac{4}{\mathcal{V}}\sum_{\bf{k}}
	\frac{a_{\bf{k+q}}^- a_\vec{k}^-
	(E_{\bf{k+q}}^- - E_\vec{k}^-) f(E_\vec{k}^- - \mu)}
	{\omega^2 - (E_{\bf k+q}^- - E_\vec{k}^-)^2} + O(q^3) \label{eq:chi-} \\
	\chi_\times({\bf q}, \omega) &= \frac{4q^2l^2}{\mathcal{V}}\sum_\vec{k}
	\frac{a_\vec{k}^+ a_\vec{k}^-
	(E_\vec{k}^+ - E_\vec{k}^-)f(E_\vec{k}^- - \mu)}
	{\omega^2 - (E_\vec{k}^+ - E_\vec{k}^-)^2} + O(q^3).
	\label{eq:chi-cross}
\end{align}
\end{widetext}
We use these expressions to compute $\omega_p$ at low-temperature by solving for $\epsilon(0,\omega_p)=0$ in \eqref{eq:epsilon-RPA}.

\section{Plasma frequency of doped HMA \label{sec:wp}}
If we take a parabolic form for the host CB, $E_\vec{k}~=~k^2/2m$, the small-$\vec{q}$ term in \eqref{eq:chi-} can be calculated analytically. The details of the derivation are in Appendix~\ref{sec:the-integral}, and the result is
\begin{multline}
	\frac{4\pi e^2\chi_-}{q^2}\approx \frac{\omega_{p0}^2}{\omega^2}\\
	= \frac{4 e^2 k_F^3}{3\pi m\omega^2}\left[\frac{1}{2}\left(1 - \frac{E_{k_F} - E_d}
	{\sqrt{(E_{k_F} -E_d)^2 + 4V^2x}}
	\right)\right]^3,
	\label{eq:wp0}
\end{multline}
which defines $\omega_{p0}$, the plasma frequency when neglecting the interband transitions, which manifest through $\chi_\times$.
Notice that without the cubed bracket, $\omega_{p0}$ would be the famous plasma frequency of an electron gas with Fermi momentum~$k_F$.
One can check that the bracket is in fact $a_k^-$, given in \eqref{eq:weights}, and as we discuss further below, the presence of $(a_k^-)^3$ in $\omega_{p0}^2$ causes a non-trivial modification in the scaling of $\omega_p$ in HMAs.

Next, to represent the $\chi_\times$ contribution to the dielectric function,
it is useful to define
\begin{align}
    \label{eq:epsilon-cross}
	&\epsilon_\times(\omega) = 1 - \frac{4\pi e^2 \chi_\times}{q^2}\\
	&= 1 + \frac{8e^2l^2V^2 x}{\pi}\int_0^{k_F} \frac{k^2 dk}
	{(E_k^+ - E_k^-)[(E_k^+ - E_k^-)^2 -\omega^2]},\nonumber
\end{align}
where the last equality uses the $\vec{q}\to0$ limit of \eqref{eq:chi-cross}. 
While the integral  in \eqref{eq:epsilon-cross} cannot be evaluated in closed form, we see that it diverges if $\omega \geq E_\times$, where $E_\times$ is defined in \eqref{eq:E-cross}.
Due to this divergence, the $\omega$-dependence of $\epsilon_\times(\omega)$ is strongest when $\omega$ is close to $E_\times$, and numerical evaluation shows that for smaller $\omega$, $\epsilon_\times$ is only weakly dependent on $\omega$.

Setting $\epsilon=0$ in \eqref{eq:epsilon-RPA} and using Eqs.~(\ref{eq:wp0},\ref{eq:epsilon-cross}), we find
\begin{equation}
	\omega_p^2 = \frac{\omega_{p0}^2}{\epsilon_\times(\omega_p)}.
	\label{eq:the-equation}
\end{equation}

\begin{figure}[t]
	\includegraphics[width=\columnwidth]{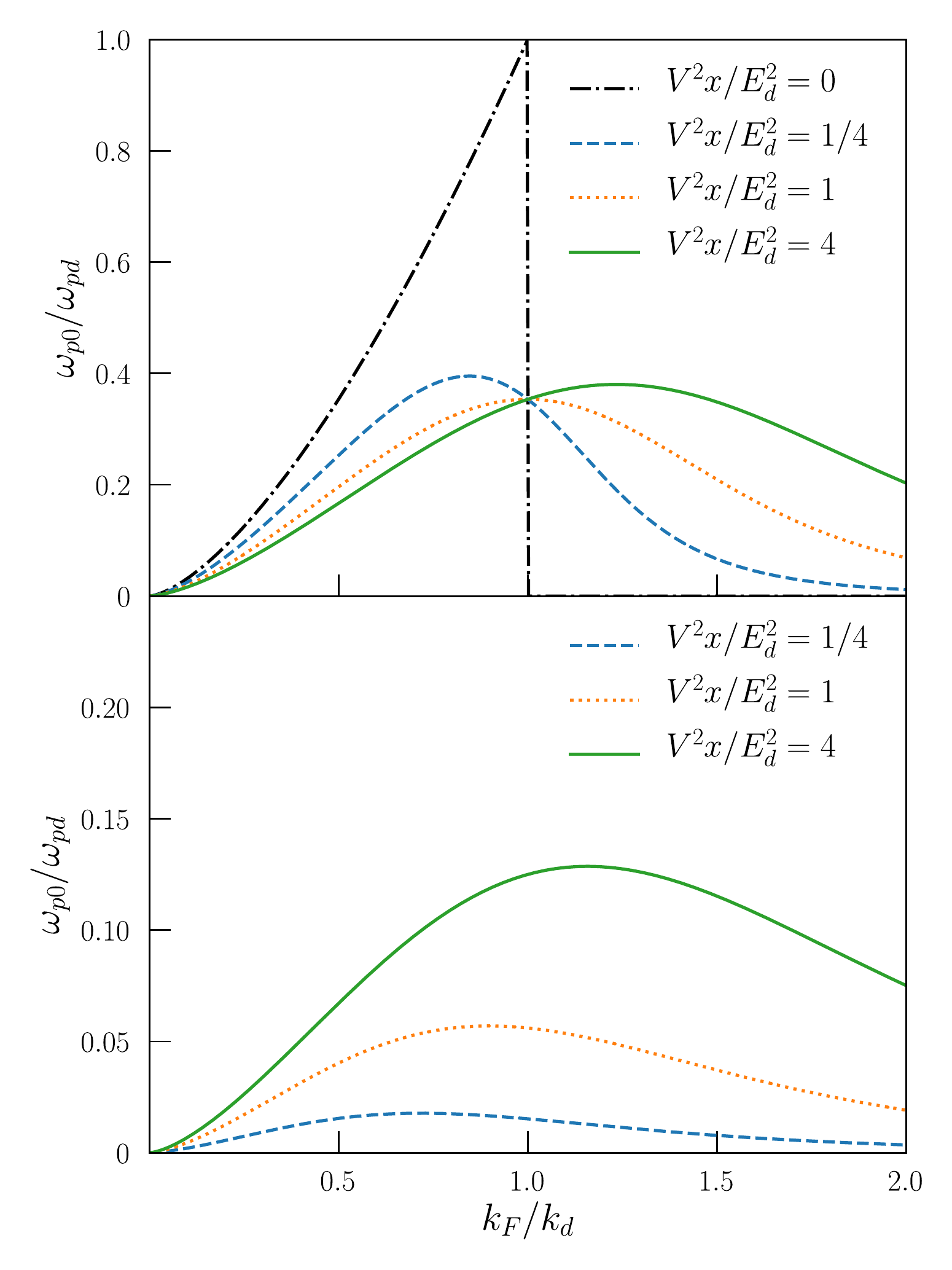}
	\caption{Intraband plasma frequency $\omega_{p0}(k_F)$ of $\Em$ band, according to \eqref{eq:wp0} for selected values of $V^2x/E_d^2$.
	The top panel has $E_d > 0$, and the bottom has $E_d < 0$. Increasing $x$ increases $\omega_{p0}$ when $E_d<0$ and for heavily doped bands with $E_d>0$, $k_F>k_d$. Contrastingly, increasing $x$ in lightly doped bands ($k_F<k_d$) with $E_d>0$ decreases $\omega_{p0}$.
}
	\label{fig:wp0}
\end{figure}

We now show how $\omega_p$ and $\omega_{p0}$ change with alloy fraction $x$ and doping, parameterized by $k_F$. Figure~\ref{fig:wp0} shows $\omega_{p0}$ against $k_F$ for selected values of $V^2x$. We normalize the $k_F$~axis using the natural inverse length scale, $k_d$, defined in \eqref{eq:kd}.
The $\omega_{p0}$~axis is normalized to $\omega_{pd}$, which is the plasma frequency of a free electron gas with effective mass $m$ when the Fermi momentum is equal to $k_d$; that is, $\omega_{pd} = \sqrt{4e^2k_d^3/3\pi m}$.

When $E_d > 0$ (Fig.~\ref{fig:wp0} top), $\omega_{p0}$ decreases with  $x$ if $k_F < k_d$, while for higher filling $\omega_{p0}$ increases with  $x$. For $k<k_d$, the lower band largely has the propagating character of the unperturbed CB, while for $k>k_d$, it is mostly made from the localized impurities, as seen in Fig.~\ref{fig:bands-weights-p}b; this crossover causes the change in the behavior of $\omega_{p0}$ with $k_F$.

When $E_d < 0$ (Fig.~\ref{fig:wp0} bottom), the $\Em$  band has largely localized impurity state character for all $k$, as shown in Fig.~\ref{fig:bands-weights-n}b. In this case $\omega_{p0}$ increases with $x$ for all levels of filling. Note that in this case $\omega_{p0}$ is significantly smaller than $\omega_{pd}$ even for relatively large values of $V^2x$; the lower share of propagating states in $\Em$ reduces the associated plasma frequency.

The plasma frequency including interband effects is given by the solution of \eqref{eq:the-equation}, which can be found numerically. However, since the $\omega$-dependence of $\epsilon_\times(\omega)$ is only strong near $E_\times$,
for $\omega_{p0}$ sufficiently smaller than $E_\times$ we can approximate the solution by, $\omega_p \approx \omega_{p0}/\sqrt{\epsilon_\times(0)}$. But if $\omega_{p0}$ is near or larger than $E_\times$, then the diverging $\epsilon_\times(\omega)$ keeps the solution below $E_\times$.
Therefore, $\omega_p$ is always smaller than both $\omega_{p0}$ and $E_\times$.

\begin{figure}[t]
	\includegraphics[width=\columnwidth]{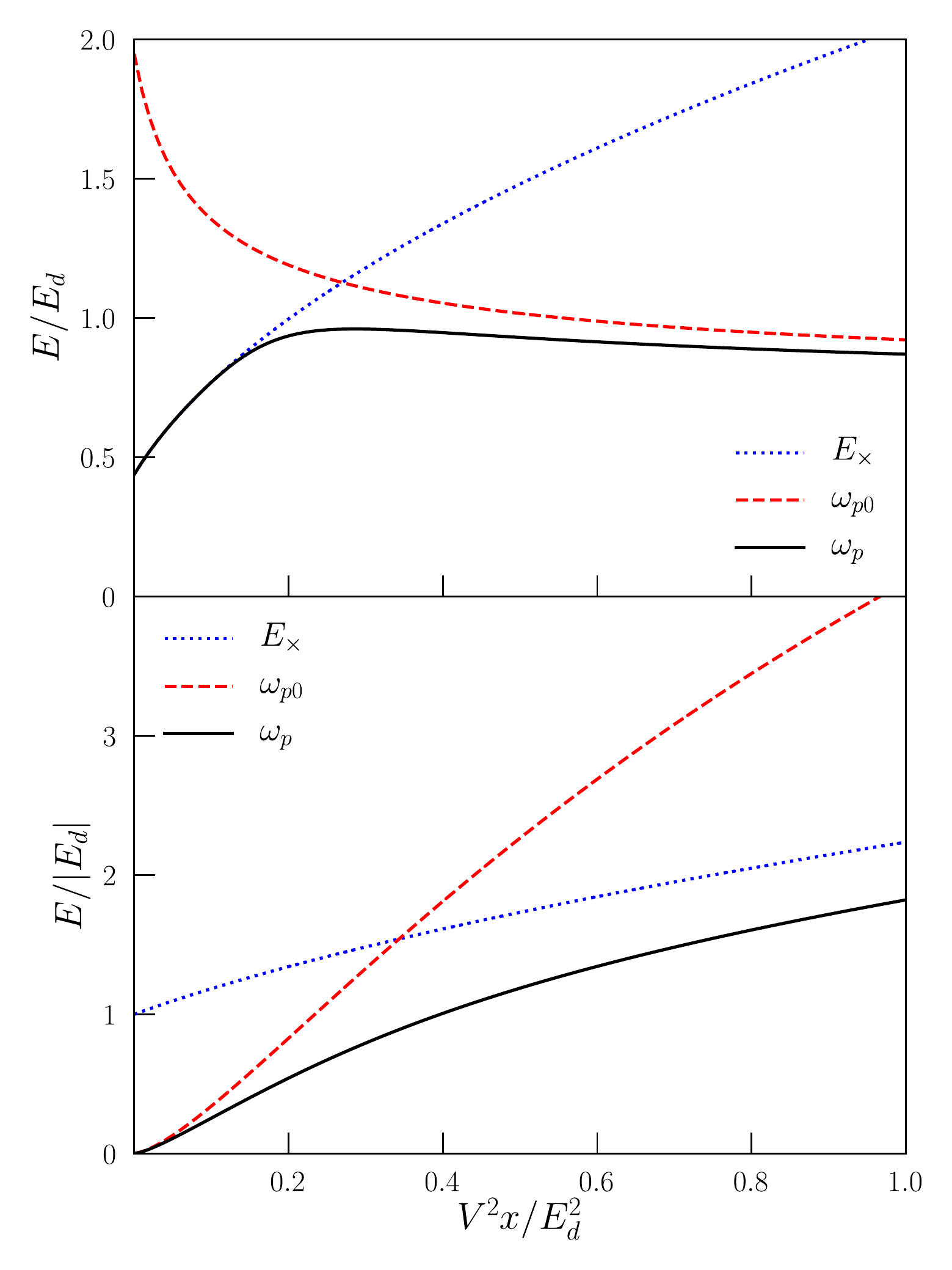}
	\caption{Plasma frequency $\omega_p$ (solid) from numerical solution of \eqref{eq:the-equation} as a function of $V^2x/E_d^2$. The energy axis is normalized to $|E_d|$.
	$\omega_p$ is bounded below the intraband plasma frequency $\omega_{p0}$ (dashed), defined in \eqref{eq:wp0}, and the smallest interband gap $E_\times$ (dotted), defined in \eqref{eq:E-cross}.
	The filling factor is $k_F = 3k_d/4$.
	Top panel has $E_d > 0$ with $\omega_{pd} = 3E_d$ and $l = 1/2k_d$.
	Bottom panel has $E_d < 0$ with $\omega_{pd} = 75 |E_d|$, and $l = 1/10k_d$.
	\label{fig:wp}}
\end{figure}

To show how $\omega_p$ is bounded by $\omega_{p0}$ and $E_\times$, 
Figure~\ref{fig:wp} plots two example solutions of \eqref{eq:the-equation} against $V^2x/E_d^2$. The top panel has $E_d > 0$ and the bottom has $E_d < 0$. The filling factor is taken to be the same in both cases, $k_F = 3k_d/4$.
In order to make the bounding effects clearly visible, a rather large value for $\omega_{pd}$ has been chosen in the case of $E_d < 0$, corresponding to a case with $E_d$ just below the CB minimum.
For $E_d > 0$ (top), $E_\times$ bounds $\omega_p$ for small $x$, while for $E_d < 0$ (bottom), it bounds $\omega_p$ for larger $x$.

The differences between plasmons in HMAs and free electron gases can be seen in the low density limit, when one might expect recovery of the free-electron result with a modified effective mass.
In the low-density limit, the filled part of the $\Em$ band can be approximated by a parabolic band of effective mass $m_-$, given by
\begin{equation}
	\frac{m}{m_-} = \frac{1}{2}\left(1 + 
	\frac{E_d}{\sqrt{E_d^2 + 4V^2 x}}\right).
	\label{eq:m-}
\end{equation}
Also, for $k \ll k_d$ we find that $a_k^-$ in \eqref{eq:weights} becomes approximately independent of $k$ and approaches $m/m_-$, which we also call $a_-$.
Moreover, the electron density in the HMA is
$n = \int_0^{k_F}d\vec{k}a_\vec{k}^-/4\pi^3$, which means in this limit,
$n~\approx~a_- k_F^3/3\pi^2$.

Then if $\omega_p$ had the standard scaling $\sqrt{n/m^*}$,  it would scale as $a_-$, since $m_- \propto 1/a_-$ and $n\propto a_-$.
However, $\omega_p$ in fact scales as $a_-^{3/2}$, which follows immediately from \eqref{eq:wp0}, where the quantity in brackets is $a_-$, as discussed previously. Note that in the low-doping regime, $\omega_p\approx\omega_{p0}$ because $\omega_{p0} \ll E_\times$ and $k_F$ is small enough that $\epsilon_\times \approx 1$. 

This anomalous scaling can be observed in a set of HMAs with varying $x$ (which changes $a_-$ or equivalently $m_-$) in which doping ensures that $n$ is held constant. If $\omega_p$ with fixed $n$ obeyed the standard free-electron effective mass result,  then as $x$ changes, $\omega_p$ would scale as $1/\sqrt{m_-}\sim \sqrt{a_-}$. Our theory instead predicts that in this fixed-$n$ case, $\omega_p$ in fact scales with $1/m_-\sim a_-$.

The unexpected extra factor of $\sqrt{a_-}$ in our result is due to the quadratic contribution of the weighting factor, $a_\vec{k}^-$, in the density susceptibility. This non-trivial scaling feature can be seen as a signature that reveals the density-density response mechanism that underlies the collective mode of plasma oscillations.
Therefore, the special state distribution in HMAs carries a qualitative effect on their bulk plasma frequency all the way to the low-density limit.
Since low levels of doping are most likely to be achievable in HMAs, we expect this peculiar scaling to be the most prominent prediction of our model to be checked by experiments.

\section{Experimental Signatures \label{sec:conclusion}}

In typical HMAs, in particular when the localized state couples to the CB of the host, the impurity level often falls within a few hundred meV below or above the CB edge.
Considering that $V$ is generally of the order of a few eV, and with the typical light effective masses of the host CB in III-V and II-VI materials, $\hbar\omega_p$ can be as large as a few hundred meV.
This scale suggests that a possible plasmonic material made by doping HMAs would operate in the range of mid-infrared or lower frequencies.
Based on a limited review of the literature
\cite{walukiewicz2008-book-chapter, adachi-ch6, bac-original, heyman-17, bac-ii-vi, kudrawiec-GaPNAs, bac-tb-spds, oreilly-bac-tb-kp, Seifikar_2014, bac-GaNSb, bac-param-1, bac-param-2, bac-param-3, bac-param-4, bac-param-5, bac-param-6, bac-param-7},
Table~\ref{tab:param-range} shows the typical range of the fixed parameters of the BAC model for III-V and II-VI HMAs with CB anti-crossings.

\begin{table}[ht]
	\caption{Typical range of fixed parameters of BAC model for HMAs, where $m_e$ is the free electron mass. \label{tab:param-range}}
	\begin{ruledtabular}
		\begin{tabular}{ccc}
		 $E_d$ [eV] & $V$ [eV] & $m$ [$m_e$]\\
		\hline
		-0.6 -- 0.4 & 1 -- 3 & 0.02 -- 0.15
		\end{tabular}
	\end{ruledtabular}
\end{table}

To examine a realistic and flexible case, we consider the quaternary alloy GaN$_x$P$_y$As$_{1-y-x}$. For this HMA we consider the nitrogen atoms to provide the localized states in a GaPAs host material. By varying the concentration of phosphorus, both positive and negative $E_d$ are realizable~\cite{kudrawiec-GaPNAs}. Transient absorption studies have been performed  on two realizations of this alloy~\cite{heyman-17}, which allows us to extract an estimate of the matrix element $l$  (see Appendix~\ref{sec:matrix-element}).
From those results, we estimate $l$ to be between 8 and 11 $\AA$, and pick $l = 10$~$\AA$ for the following calculations.
For the numerical values of $E_d$ and $V$, we rely on Ref.~\onlinecite{kudrawiec-GaPNAs}, and for the effective masses and the host's energy gaps we use Refs.~\onlinecite{adachi-ch6,band-param-III-V}.
For $V$ and $m$, we assume a linear interpolation with $y$. But for $E_d$, relying on Ref.~\onlinecite{band-param-III-V}, we also take into account bowing,
$E_d(y) = (1-y)E_d\vert_{y = 0} + yE_d\vert_{y = 1}+ y(1-y)C$, with $C = 0.19$ eV.
These parameters are listed in Table~\ref{tab:GaNPAs}.

\begin{table}[ht]
	\caption{Parameters of BAC model for GaN$_x$P$_y$As$_{1-y-x}$
	\cite{kudrawiec-GaPNAs,adachi-ch6,band-param-III-V}, where $m_e$ is the free electron mass.
}
	\label{tab:GaNPAs}
	\begin{ruledtabular}
		\begin{tabular}{lcc}
		Parameters & $y = 0$ & $y = 1$\\
		\hline
		$E_d$ [eV] & 0.22 & -0.6\\
		$V$ [eV] & 2.8 & 3.05\\
		$m$ [$m_e$] & 0.067 & 0.13\\
		\end{tabular}
	\end{ruledtabular}
\end{table}

\begin{figure}[ht]
	\includegraphics[width=\columnwidth]{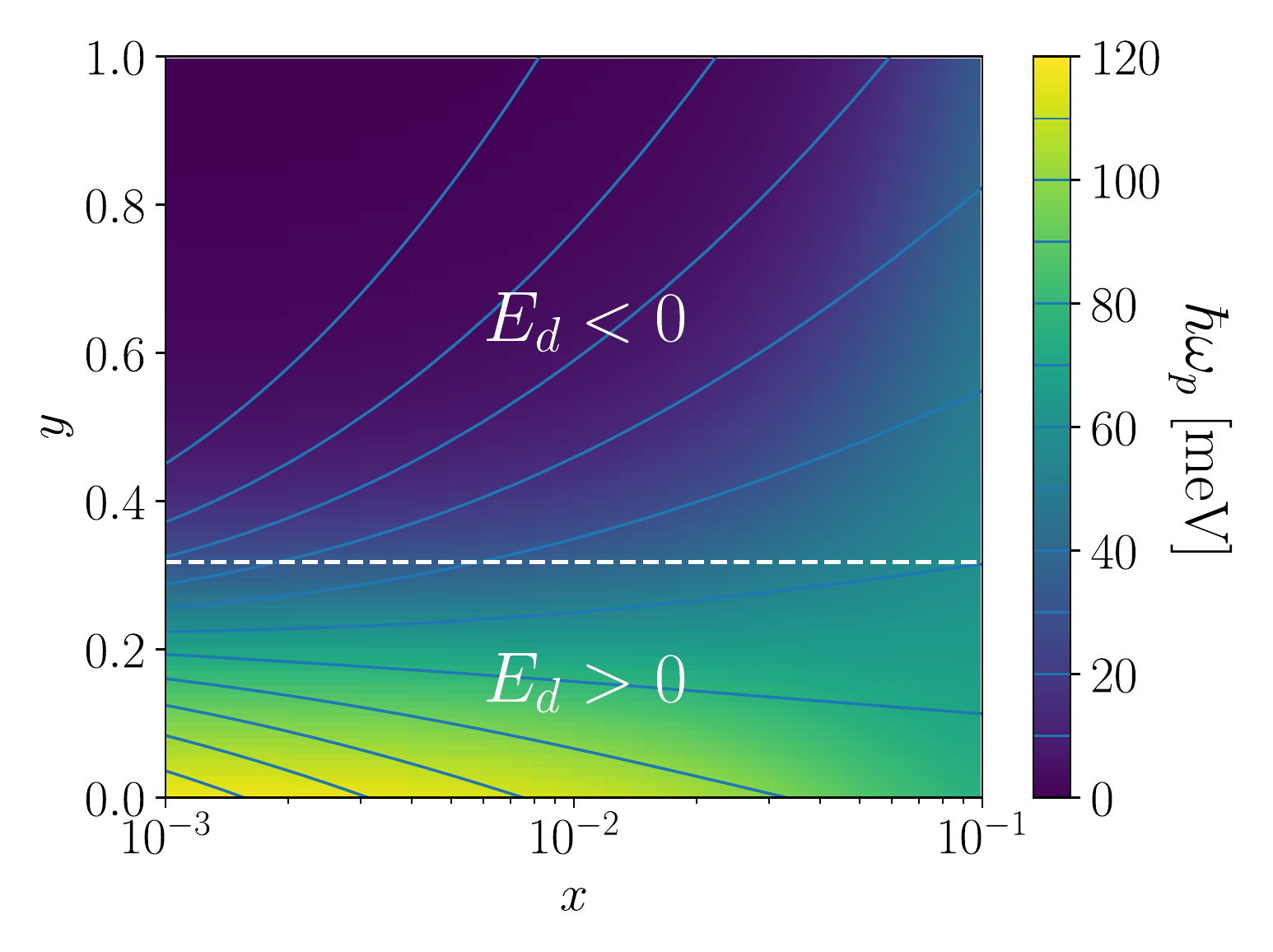}
	\caption{Plasmon energy, $\hbar\omega_p$, for GaN$_x$P$_y$As$_{1-y-x}$ at fixed electron density, $n = 10^{18}$~cm$^{-3}$.
	BAC parameters listed in Table~\ref{tab:GaNPAs}, and explained in the paragraph above it, are used for GaN$_x$P$_y$As$_{1-y-x}$, and a fixed matrix element, $l = 10 \AA$, is assumed for the calculations.
	Plasma frequencies are highest below the dashed line, where $E_d$ is positive.}
	\label{fig:GaPAsN}
\end{figure}

We consider moderately doped materials with fixed $n = 10^{18}$~cm$^{-3}$ and calculate $\omega_p$ by numerically solving \eqref{eq:the-equation}, with results in Fig.~\ref{fig:GaPAsN} for all $y$ and $10^{-3}~<~x~<~10^{-1}$.
At this doping, $\hbar\omega_p$ is largest near the GaAs limit at small $y$ (i.e., positive $E_d$) and small $x$, approaching 120~meV. 
Here, $E_\times$ is always larger than 180~meV and hence doesn't have a significant bounding effect on $\omega_p$.
The phosphorus fraction $y$ at which $E_d$ changes sign is marked on the plot. One can see that $\omega_p$ is generally smaller on the $E_d < 0$ side, where the $\Em$ band is predominantly made of the localized levels, and $a_-$ is generally small, similar to the results in Fig.\ \ref{fig:wp0}.

Our simple two-band model for the dielectric function takes $\epsilon_\infty$ to be 1, which is not correct for most real materials. Based on that consideration, one would expect to find smaller $\omega_p$ in an experiment than what we report here. 
 The predictive nature of our model is in how $\omega_p$ changes upon changing of $x$ and doping, not the quantitative values.

Given that the plasmon resonance is expected to appear in mid-infrared range, dielectric permittivity measurements such as ellipsometry \cite{measure-ellips-TiN} or Fourier-transform infrared  spectroscopy \cite{measure-ftir-ITO,measure-ftir-InP}  are viable candidates to extract $\omega_p$ in these materials. If the HMA supports plasmonic propagating modes, then indirect techniques such as the proposal in Ref.~\onlinecite{measure-thz-indirect} can also give information on  $\omega_p$.

The constraint that $E_\times$ imposes on $\omega_p$ could be important in certain cases. One would expect that effect to appear when $x$ is low, especially when $E_d$ is close to 0.

It is possible  that $\omega_p$ for these materials may be close to their optical phonon resonance, especially in the low doping regimes and when $E_d<0$, when $\omega_p$ is small, as the typical frequency range of optical phonons is in THz.
In such a case, one needs to be careful when extracting $\omega_p$ from permittivity measurements.

The great tunability of HMAs through both doping and alloy fraction allows for optimization of their potential plasmonic applications.
In cases such as the quaternary GaN$_x$P$_y$As$_{1-y-x}$, the relative location of the localized level can also be tuned, providing even more handles for tuning. 
Moreover, the gap between the $\Em$ and the $\Ep$ bands could potentially allow for tuning the plasma frequency in a minimal loss regime~\cite{lossless-matal}; a fact that suggests the potential use of some well-tuned doped HMAs as low-loss plasmonic materials in the mid-infrared region. 

The calculations presented in this work assume infinitesimal broadening factor $\Gamma$ and ignore the disorder effects of the random alloy, which can allow violation of momentum conservation. Future work exploring the implications of these effects for plasmons may be important. 

\begin{acknowledgments}
We acknowledge funding from the NSERC CREATE TOP-SET program, Award Number 497981.
\end{acknowledgments}

\appendix

\section{Long-wavelength limit of the~matrix~elements \label{sec:matrix-element}}
Tight binding (TB) models have been used to describe HMAs band structure~\cite{bac-tb-spds,oreilly-bac-tb-kp,kudrawiec-GaPNAs}.
Here we use a TB model combined with experimental results to justify the long-wavelength approximation of the matrix element $|\langle \vec{k}, - |e^{i{\bf q\cdot r}}| {\bf k'}, +\rangle|^2$ in \eqref{eq:cross-me}, including momentum conservation.
In a TB model, wavefunctions in band $i$ are given by~\cite{marder-tight-binding}
\begin{equation}
	|\vec{k}, i\rangle = \frac{1}{\sqrt{N}}\sum_{\bf R}e^{i{\bf k\cdot R}}|\phi_\vec{k}^{(i)}({\bf r -R})\rangle,
	\label{eq:tb-wf}
\end{equation}
where $\bf k$ is the crystal momentum, $\bf r$ is the position in the real space, and the sum is over all lattice points $\bf R$.
$N$ is a normalization factor, and if we neglect overlaps between different sites,
is equal to the number of lattice points, while $\phi_\vec{k}^{(i)}$'s are an orthonormal set of orbital wavefunctions given by linear combinations of atomic (or molecular) orbitals of each unit cell.
Using the form in~\eqref{eq:tb-wf} we have
\begin{align}
	\label{eq:me-exact}
	&\langle \vec{k}', i |e^{i{\bf q\cdot r}}|\vec{k}, j \rangle = \\
	&\frac{1}{N}\sum_{{\bf R,R'}}
	e^{i({\bf k\cdot R- k'\cdot R'})}
	\langle \phi_{\bf k'}^{(i)}({\bf r-R'})|e^{i{\bf q\cdot r}}|\phi_\vec{k}^{(j)}(\bf {r-R})\rangle.
	\nonumber
\end{align}
If we neglect the off-diagonal terms, ${\bf R} \neq {\bf R}'$,
then multiplying and dividing \eqref{eq:me-exact} by $e^{i\bf{q\cdot R}}$, and shifting  ${\bf r - R}\to {\bf r}$ in the matrix elements, we get
\begin{align}
	\label{eq:me-aprox}
	\langle \vec{k}', i & |e^{i{\bf q\cdot r}}|\vec{k}, j \rangle \\
	\approx&\langle \phi_{\bf k'}^{(i)}({\bf r})|e^{i{\bf q\cdot r}}|\phi_\vec{k}^{(j)}({\bf r})\rangle
	\frac{1}{N}\sum_{\bf R}
	e^{i({\bf k + q- k'})\cdot {\bf R}} \nonumber\\
	=&\langle \phi_{\bf k'}^{(i)}({\bf r})|e^{i{\bf q\cdot r}}|\phi_\vec{k}^{(j)}(\bf {r})\rangle
	\delta_{\bf{k', k+q}} \nonumber,
\end{align}
which immediately implies momentum conservation.

In the long-wavelength limit, we have $e^{i{\bf q\cdot r}} \approx 1 + i{\bf q\cdot r}$,
and $\vec{k}' \approx \vec{k}$, due to momentum conservation.
Therefore, the orthonormality of $\phi_\vec{k}^{(i)}$'s implies that the intraband matrix element is  
$|\langle \vec{k}', i |e^{i{\bf q\cdot r}}|\vec{k}, i \rangle|^2\approx \delta_{\bf{k', k+q}}$, while the interband matrix elements are
\begin{equation}
	\langle \vec{k}', i |e^{i{\bf q\cdot r}}|\vec{k}, j \rangle \approx
	i{\bf q}\cdot\langle \phi_\vec{k'}^{(i)}({\bf r})|{\bf r}|\phi_\vec{k}^{(j)}(\bf {r})\rangle
	\delta_{\bf{k', k+q}},
	\label{eq:inter-me-approx}
\end{equation}
as the first term vanishes due to orthogonality of $\phi_\vec{k}^{(i)}$'s.
If a TB model with localized orbitals is a suitable model for describing band structure of HMAs, then \eqref{eq:inter-me-approx} justifies the form that we choose in~\eqref{eq:cross-me}, suggesting that $l$ must be a length scale of the order of the lattice parameter.

The interband matrix element in \eqref{eq:cross-me} is related to the interband absorption, which permits an independent estimate of its magnitude.
If $|s\rangle$
is an eigenstate of a system with energy $E_s$, the absorption coefficient is \cite{marder-kubo-greenwood} 
\begin{equation}
	\alpha(E) = \frac{4\pi^2e^2}{\overline{n}\hbar c E \mathcal{V}}\sum_{ss'}
	\left|\left\langle s \left| \frac{\hbar \hat{\vec{P}}}{m_e}\right| s' \right\rangle\right|^2
	\delta(E_{s'} - E_s -E)(f_s - f_{s'}),
	\label{eq:general-alpha}
\end{equation}
where $\overline{n}$ is the refractive index and $\hat{\vec{P}}$ is the momentum operator.

It is also straightforward to see that if the only momentum dependence of a single particle Hamiltonian is the kinetic part, ${\bf P}^2/2m_e$, as is the case for Hamiltonians describing band structures of crystals, then we have
\begin{equation}
	(E_{s'} - E_s)\langle s|\vec{r}| s' \rangle = \frac{i\hbar}{m_e}\langle s|\vec{P}| s' \rangle,
	\label{eq:P-r-connection}
\end{equation}
where $\vec{r}$ is the position operator.
Equation (\ref{eq:P-r-connection}) shows how the position matrix element, as in \eqref{eq:inter-me-approx}, is related to the momentum matrix element that is present in the expression of $\alpha$, \eqref{eq:general-alpha}. 

Assuming that $\hat{\vec{P}}$ also conserves momentum, the sum in \eqref{eq:general-alpha} reduces to a single sum over $\vec{k}$.
Now consider that an $|s\rangle$,$|s'\rangle$ pair in \eqref{eq:general-alpha} are $|\vec{k}, \pm\rangle$, two eigenstates of the Hamiltonian in the $\Ep$ and $\Em$ bands, respectively.
Then the length scale $l$ is given by $|\langle \vec{k}, + | \vec{r} | \vec{k}, - \rangle|^2 = l^2$. Equation (\ref{eq:P-r-connection}) then allows $|\langle s|\hbar\vec{P}/m_e | s' \rangle|^2$ in \eqref{eq:general-alpha} to be written as $l^2E^2$, since the delta function enforces $E_\vec{k}^+ - E_\vec{k}^- = E$. 

With these considerations, and approximating that $l$ is independent of $\vec{k}$, for $\Em \to \Ep$ absorption coefficient we can write
\begin{equation}
	\alpha(E) = \frac{4\pi^2 e^2 E l^2}{\overline{n}\hbar c}D_j(E),
	\label{eq:alpha-Epm}
\end{equation}
where $D_j(E)$ is the $\bf k$-conserving joint density of states with energy  $E$.

In the transient absorption measurements of Ref.~\onlinecite{heyman-17}, photoexcitation populates the $\Em$ levels, and a probe beam is used to determine the absorption from these transiently populated states. We consider sample S205B, whose transient absorption spectrum 2~ps after photoexcitation is shown in their Fig.~3a.
We consider the absorption at $E = 0.8$~eV. Assuming that the absorption change is entirely due to the $\Em \to \Ep$ transitions, and noting that the thickness of the absorptive layer is 0.5~$\mu$m, we estimate $\alpha(0.8\text{ eV}) \approx 700$ cm$^{-1}$.

Then, to use \eqref{eq:alpha-Epm} to estimate $l$, we need to estimate $D_j(E)$. First we calculate $D_j(E)$ between the full bands $\Em$ and $\Ep$ while taking the weighting factors, $a_\pm$, into account. We then need to reduce $D_j(E)$ to account for the partial occupancy of $\Em$. Since the transient absorption experiment considers excitation from a photoexcited population in $\Em$ that is not in equilibrium, and since the $\Em$ bandwidth is not very wide, we consider the electrons to be uniformly distributed in $\Em$. We then reduce $D_j(E)$ by the ratio $n_-/n_{\rm max}$, where $n_-$ is the electron density in $\Em$, reported at 2~ps in Fig.~5 in Ref.~\onlinecite{heyman-17} to be approximately  $2\times 10^{18}$~cm$^{-3}$.
For $n_{\rm max}$, we consider two limiting approximations: first, $n_\text{max}\approx 5.2\times 10^{19}$~cm$^{-3}$ is the concentration of electrons in a completely filled $\Em$; second, 
$n_\text{max}\approx 3.2 \times 10^{19}$~cm$^{-3}$ is the concentration of electrons in the $\Em$ band if it is filled up to the point where $E = 1.6$~eV (the upper limit of the observed $\Em \to \Ep$ absorption band) is accessible.
Using $\overline{n} = 3.25$ \cite{n-GaAsP},
these approximations produce a range for $l$ between 8 and 11 $\AA$.
Since in the particular case we are considering, interband transitions don't have a significant effect on plasma frequency, $\omega_p$ changes by at most about $3\%$, as $l$ varies between 8 and 11 $\AA$.
We use $l = 10$~$\AA$ in Fig.~\ref{fig:GaPAsN}.

\section{Analytic calculation of $\omega_{p0}$ \label{sec:the-integral}}
In order to find $\omega_{p0}$ in \eqref{eq:wp0}, one needs to expand the $\chi_-$-integral in \eqref{eq:chi-} up to the second order, for a parabolic conduction band, $E_k = k^2/2m$. Performing the angular part of the integral, the first term in the expansion vanishes, and the next terms, which are proportional to $q^2$, have two parts
\begin{align}
	\label{eq:chi2-int}
	&\chi_-^{(2)}(q,\omega) = \frac{q^2}{8\pi^2 m \omega^2} \\
	\times&\left\lbrace\int_0^{k_F}k^2
	\frac{\left[E_d - E_k + \sqrt{(E_k - E_d)^2 + 4V^2x}\right]^3}{\left[(E_k - E_d)^2 + 4V^2x\right]^{3/2}}dk
	\right.\nonumber\\
	&-\frac{4V^2x}{m}\left.\int_0^{k_F}k^4
	\frac{\left[E_d - E_k + \sqrt{(E_k - E_d)^2 + 4V^2x}\right]^2}{\left[(E_k - E_d)^2 + 4V^2x\right]^{5/2}}dk
	\right\rbrace \nonumber\\
	& = \frac{q^2}{8\pi^2 m \omega^2}\left(I_1 - I_2\right), \nonumber
\end{align}
where the last equation defines the integrals $I_1$ and $I_2$.  Since $E_k = k^2/2m$, the change of variables $u = k^2$ allows integrating $I_1$ by parts to obtain
\begin{equation}
	I_1 = \frac{k_F^3}{3}\left[1 - 
	\frac{E_{k_F} - E_d}{\sqrt{(E_{k_F}-E_d)^2 + 4V^2x}}\right]^3 + I_2.
	\label{eq:I1-bp}
\end{equation}
Using this result in  \eqref{eq:chi2-int} gives \eqref{eq:wp0}, which defines $\omega_{p0}$.

\bibliography{References}

\end{document}